# Exact Solutions for Domain Walls in Coupled Complex Ginzburg – Landau Equations

Tat Leung YEE[(1)], Alan Cheng Hou TSANG[(2)],

Boris MALOMED[(3)], and Kwok Wing CHOW[*(2)]

(1) = Department of Mathematics and Information Technology,

Hong Kong Institute of Education, Tai Po, New Territories, Hong Kong

(2) = Department of Mechanical Engineering, University of Hong Kong,

Pokfulam, Hong Kong

(3) = Department of Physical Electronics, School of Electrical Engineering,

Tel Aviv University, Tel Aviv 69978, Israel

* = Corresponding author

Email: kwchow@hku.hk       Fax: (852) 2858 5415

# ABSTRACT

The complex Ginzburg Landau equation (CGLE) is a ubiquitous model for the evolution of slowly varying wave packets in nonlinear dissipative media. A front (shock) is a transient layer between a plane-wave state and a zero background. We report exact solutions for *domain wall*s, i.e., pairs of fronts with opposite polarities, in a system of two coupled CGLEs, which describe transient layers between semi-infinite domains occupied by each component in the absence of the other one. For this purpose, a modified Hirota bilinear operator, first proposed by Bekki and Nozaki, is employed. A novel factorization procedure is applied to reduce the intermediate calculations considerably. The ensuing system of equations for the amplitudes and frequencies is solved by means of computer-assisted algebra. Exact solutions for mutually-locked front pairs of opposite polarities, with one or several free parameters, are thus generated. The signs of the cubic gain/loss, linear amplification/attenuation, and velocity of the coupled-front complex can be adjusted in a variety of configurations. Numerical simulations are performed to study the stability properties of such fronts.

# 1. Introduction

Weakly nonlinear waves in spatially extended nonlinear dissipative systems frequently obey several ubiquitous evolution models, a well-known example being the complex Ginzburg-Landau equation (CGLE) for the slowly varying amplitude $A$ of the wave,[1-8]

$$iA_t + pA_{xx} + q|A|^2 A = i\gamma A , \qquad (1)$$

with complex dispersion and nonlinearity coefficients $p$ and $q$, and real linear gain coefficient $\gamma$. Imaginary parts of $p$ and $q$, with proper signs, account for the diffusive and nonlinear losses respectively. Another genetic model is represented by a system of nonlinearly coupled CGLEs, see eqs. (5) and (6) below.

The CGLE is not integrable, and hence the powerful tools associated with its conservative counterpart, the nonlinear Schrödinger equation (NLSE), are not applicable. Nevertheless, several techniques which produce analytical solutions for solitary pulses have been developed. Unlike the celebrated solitons, where only dispersive and nonlinear effects need to be mutually balanced, solitary pulses in the CGLE must in addition maintain the equilibrium between energy gain and loss. Consequently, solitary pulses are usually represented by isolated exact solutions of the CGLE and related equations,[9] rather than continuous families, with the latter more typical for solitons in the NLSE.

The Hirota bilinear method is a well established method for obtaining multi-soliton expressions in integrable nonlinear evolution equations.[10] To extend the usage of this method to the CGLE, a modified bilinear operator, pioneered by Bekki and Nozaki, is needed.[11] The aim of the present work is to develop special techniques, in conjunction with the Bekki-Nozaki operator, with the aim of producing exact solutions for the system of coupled CGLEs, see eqs. (5) and (6) below.

Our analysis deals with front solutions, also known as 'domain walls', 'kinks' and 'shocks' in other contexts. A front is a sharp transition between a plane-wave state and a zero background in the asymptotic fields. Front solutions have been previously investigated in several settings, including those based on coupled equations. In particular, the interaction of fronts was studied in a system consisting of a real Ginzburg-Landau equation coupled to a mean field.[12] Sometimes the term 'front' refers to 'phase fronts', which separate domains of different phase-locked states. In that case, the Benjamin-Feir (modulation) instabilities may lead to explosion of the front.[13] Fronts propagating into an unstable medium were considered too.[14, 15] The transition from localized pulses to fronts was studied in a CGLE with a combined cubic-quintic nonlinearity.[16]

Substantial progress has been achieved in obtaining exact solutions for systems in which one equation is linear.[17−20] In this work, we consider a fully nonlinear system of coupled CGLEs. Such nonlinear systems are relevant in many applications in hydrodynamics,[21-24] optics,[25] oscillatory media,[26] and plasma physics.[27]

It is instructive to first consider the simplest version of the system arising in thermal convection. Here nonlinearly coupled *real* Ginzburg-Landau equations for the local amplitudes, *A* and *B*, of two interacting families of static rolls with different orientations, are governed by[22-24]

$$A_t = A_{xx} - (|A|^2 + g_0|B|^2)A + A, \qquad (2a)$$

$$B_t = B_{xx} - (|B|^2 + g_0|A|^2)B + B, \qquad (2b)$$

where $g_0 > 0$ is a real coefficient accounting for the interaction. An exact solution for domain walls for this system is available solely for $g_0 = 3$:

$$A = \frac{1}{2}\left[1 + \tanh\left(\frac{x}{\sqrt{2}}\right)\right], \qquad B = \frac{1}{2}\left[1 - \tanh\left(\frac{x}{\sqrt{2}}\right)\right]. \qquad (2c)$$

For $g_0$ close to unity, an approximate solution could be found analytically.[22]

Coupled *complex* Ginzburg-Landau equations will describe the interaction of counter-propagating waves in the convection in binary fluids.[21] In fact, in addition to terms written below for eqs. (5) and (6), those equations may also include terms denoting the presence of opposite group velocities. Domain walls exist in the latter case too. Actually, they represent sources or sinks of two families of waves traveling in the opposite directions.[23] Solutions for such domain walls between traveling waves were found in an approximate form too.[23, 24]

Our goal in this work is to report on several families of *exact* solutions for domain walls in systems of two coupled CGLEs, established analytically for the first time. These new solutions are obtained by a novel factorization procedure, which will reduce the algebraic manipulations involved in the intermediate calculations considerably.

The paper is structured as follows. The modified Hirota operator pioneered by Bekki and Nozaki is reviewed, and the nonlinear model is introduced in Section 2. Reductions from the corresponding 'trilinear' to 'bilinear' equations are presented in Section 3. The new families of solutions are produced in Sections 4 and 5. Numerical simulations are performed to investigate the stability properties of the fronts (Section 6), and conclusions are drawn in Section 7.

## 2. The modified Hirota operator and the coupled CGL model

(A) *The modified Hirota operator*

The generalized Hirota operator, introduced by Bekki and Nozaki,[11, 28] is defined by

$$D_{z,x}^{M}(G \cdot f) = \left(\frac{\partial}{\partial x} - z\frac{\partial}{\partial x'}\right)^{M} G(x) \cdot f(x') \bigg|_{x=x'}, \qquad (3)$$

where $z$ may be complex and $M$ is a positive integer. Eq. (3) with $z = 1$ reduces to the ordinary Hirota operator.[10] The ordinary operator is meant if only one subscript is used:

$$D_x \equiv D_{1,x}.$$

One can easily verify the following differentiation rule:

$$D_{m,x}^N \exp(ax) \cdot \exp(bx) = (a - mb)^N \exp((a+b)x), \tag{4}$$

for complex constants $a$, $b$, $m$ and integer $N$.

(B) *The coupled CGL model*

The subject of this work is the system of coupled CGLEs for slowly varying amplitudes $A$ and $B$,

$$iA_t + p_1 A_{xx} + (q_1|A|^2 + q_2|B|^2)A = i\gamma_1 A, \tag{5}$$

$$iB_t + p_2 B_{xx} + (q_1|B|^2 + q_2|A|^2)B = i\gamma_2 B. \tag{6}$$

Complex coefficients $p_{1,2}$ and $q_1$ have the same physical meaning as the counterparts in the single component case, eq. (1), while $q_2$ accounts for the nonlinear coupling.

To apply the Hirota method, we perform the transformations

$$A = \frac{G \exp(-i\Omega_1 t)}{f^m}, \qquad B = \frac{H \exp(-i\Omega_2 t)}{f^n} \tag{7}$$

where $G$ and $H$ are complex functions, $f$ is real, while $m$ and $n$ are the following complex numbers with imaginary parts $\alpha$ and $\beta$:

$$m = 1+i\alpha, \qquad n = 1+i\beta. \tag{8}$$

The application of the modified Hirota bilinear operator (1) makes the governing model, eqs. (5) and (6), tantamount to a system of two 'trilinear' equations:

$$f[iD_{m,t} + p_1 D_{m,x}^2 + \Omega_1 - i\gamma_1 - C_1]G \cdot f$$
$$+ G\left\{-\frac{p_1 m(m+1)D_x^2 f \cdot f}{2} + C_1 f^2 + q_1 GG^* + q_2 HH^*\right\} = 0, \tag{9}$$

$$f[iD_{n,t} + p_2 D_{n,x}^2 + \Omega_2 - i\gamma_2 - C_2]H \cdot f$$
$$+ H\left\{-\frac{p_2 n(n+1)D_x^2 f \cdot f}{2} + C_2 f^2 + q_1 HH^* + q_2 GG^*\right\} = 0. \tag{10}$$

In many conventional treatment of the energy conserving NLSEs, the second terms inside the brackets in eqs. (9) and (10) are set to be zero, reducing them to bilinear equations.[29] Here we, instead, assume that these second terms will be properly factorized, and need *not* vanish. We shall restrict our attention to cases where $C_1 = C_2 = 0$ in this paper, leaving the more general case for future studies.

## 3. Fronts of opposite polarities

To look for fronts of opposite polarities for eqs. (5, 6), which interpolate between asymptotic domains carrying the plane-wave background in either component and the vanishing field in the other one, a suitable expansion scheme is

$$G = g\exp(kx - \omega t), \quad H = h, \quad f = 1 + \exp(kx - \omega t), \tag{11}$$

where $g$ and $h$ are complex constants, while $k$ and $\omega$ are real. Examining the limits of eq. (7) as $x \to \pm\infty$ will reveal that eq. (11) does represent a pair of fronts with opposite polarities. However, even with the apparently simple ansatz in eq. (11), the necessary manipulations of eqs. (9), (10) still lead to an oppressive amount of algebra.

An important simplification is to insist that the terms in the curly brackets of the trilinear eqs. (9) and (10) be factorized properly, i.e., we search for constant $\sigma_1$, $\sigma_2$ such that (for the case of $C_1 = C_2 = 0$)

$$-\frac{p_1 m(m+1) D_x^2 f \cdot f}{2} + q_1 |G|^2 + q_2 |H|^2$$

$$= q_2 |h|^2 [1 + \exp(kx - \omega t)][1 + \sigma_1 \exp(kx - \omega t)], \tag{12}$$

$$-\frac{p_2 n(n+1) D_x^2 f \cdot f}{2} + q_1 |H|^2 + q_2 |G|^2$$

$$= q_1 |h|^2 [1 + \exp(kx - \omega t)][1 + \sigma_2 \exp(kx - \omega t)]. \tag{13}$$

With these simplifications, the trilinear system now reduces to bilinear equations after dividing by the common factor $f$ as defined by eq. (11).

In this paper we restrict our attention to real $\sigma_1$, $\sigma_2$, leaving possibilities of complex values of these constants for future studies. By applying the differentiation rule (4) repeatedly, one finally arrives at the following algebraic constraints:

$$im\omega + q_2|h|^2\left[\sigma_1 - 1 - \frac{(m-2)(1+\sigma_1)}{m+1}\right] = 0, \qquad (14)$$

$$in\omega + q_1|h|^2\left[\sigma_2 - 1 - \frac{n(1+\sigma_2)}{n+1}\right] = 0, \qquad (15)$$

$$-p_1 m(m+1)k^2 = (1+\sigma_1)q_2|h|^2, \qquad (16)$$

$$-p_2 n(n+1)k^2 = (1+\sigma_2)q_1|h|^2, \qquad (17)$$

$$\sigma_1 q_2^2 = \sigma_2 q_1^2, \qquad (18)$$

and for easy reference, symbols $p_1$, $p_2$, $q_1$, $q_2$ first appeared in eqs. (5, 6); symbols $m$, $n$, in eq. (8); symbols $h$, $k$, $\omega$, in eq. (11); and $\sigma_1$, $\sigma_2$, in eqs. (12, 13). Some details of the derivation of eqs. (14 – 18) are given in the Appendix.

The other parameters, $\Omega_1$, $\Omega_2$ (angular frequencies of the envelope), $g$ (amplitude of one waveguide) and $\gamma_1$, $\gamma_2$ (linear gain/loss) in eqs. (5 – 7) are determined by further auxiliary constraints:

$$-i\omega + p_1 k^2 + \Omega_1 - i\gamma_1 + q_2|h|^2 = 0, \ \Omega_2 - i\gamma_2 + q_1|h|^2 = 0, \ |g|^2 = (\sigma_1\sigma_2)^{1/2}|h|^2, \qquad (19)$$

once the other parameters are found from eqs. (14 – 18). From the last equation in eq. (19) it follows that $\sigma_1$ and $\sigma_2$ should either be real numbers of the same sign, or complex conjugate of each other.

The actual algebraic manipulations constitute a major undertaking and are accomplished by means of a computer software. Eqs. (14 – 18) consist of five complex (or ten real) equations for *seven real* unknowns: $|h|^2$, $k$, $\alpha$, $\beta$ (see eq. (8)), $\sigma_1$, $\sigma_2$, $\omega$ and *four complex* parameters $p_1$, $p_2$, $q_1$, and $q_2$. Consequently, either

- three real constraints must be additionally imposed upon $p_1$, $p_2$, $q_1$, and $q_2$; or
- any solution of eqs. (14 – 18) can have a maximum of $7 + 4 \cdot 2 - 10 = 5$ degrees of freedom (or arbitrary parameters) in principle.

Two such families of solutions are presented in the following sections, utilizing particularly simple choices of three real constraints imposed on $p_1$, $p_2$, $q_1$, and $q_2$.

## 4. The first family of exact solutions (purely imaginary $q_1$, $q_2$)

To obtain solutions in an explicit form, we make simplifying assumptions. For the first family of exact solutions, we take

$$q_2 = q_1 = q_r + iq_i. \tag{20}$$

The first equality of eq. (20) is equivalent to two real constraints, and the third one is taken as

$$q_r = 0 \quad \text{or} \quad q_1 = q_2 = iq_i \text{ (with arbitrary } q_i\text{)}. \tag{21}$$

Various exact solutions obtained by means of the Maple software package are tabulated below.

A set of solutions with four degrees of freedom, or four arbitrary parameters $q_i$, $p_{2r}$, $p_{2i}$, $k^2$, is given by

$$p_2 = p_{2r} + ip_{2i}, \tag{22}$$

$$\beta^2 + \frac{3\beta p_{2i}}{p_{2r}} - 2 = 0, \tag{23}$$

$$\alpha^2 = 14 - \frac{9\beta p_{2i}}{p_{2r}}, \tag{24}$$

$$\sigma_2 = \sigma_1 = \frac{2\beta^2 + 5}{3}, \tag{25}$$

$$\left(\frac{4\omega}{k^2} - p_{2i}\right)^2 = 8p_{2r}^2 + 9p_{2i}^2, \tag{26}$$

$$|h|^2 = \frac{3\omega}{2q_i}, \tag{27}$$

$$p_1 = \frac{(\beta^2 + 4)[-3\alpha + (\alpha^2 - 2)i]\omega}{k^2(\alpha^2 + 1)(\alpha^2 + 4)}. \tag{28}$$

The parameters in eqs. (22 – 28) should be selected such that real solutions can be obtained, e.g., the discriminant of quadratic eq. (23) must be positive.

*As a summary, expressions (7), (8), (11), (19), and (22 – 28) yield an exact solution to the coupled CGLEs, eqs. (5, 6), if eq. (21) holds.*

The sign of $q_i$ may be arbitrary. We present the following examples:

*Example A*

As a simple particular case, we highlight the one given by eqs. (22 – 28) for $q_i > 0$ and

$$p_2 = -9 + 7i. \tag{29}$$

In this case, the exact solution is

$$p_1 = \mp \frac{5\sqrt{35}}{18} + \frac{55i}{18}, \tag{30}$$

$$\alpha = \pm\sqrt{35}, \quad \beta = 3, \tag{31}$$

$$\sigma_1 = \sigma_2 = \frac{23}{3}, \quad \omega = 10k^2, \quad |h|^2 = \frac{15k^2}{q_i}. \tag{32}$$

*Example B*

Similarly, in the particular case of eqs. (22 – 28) for $q_i < 0$, which corresponds to a cubic gain, an exact solution with $p_2$ still taken as per eq. (29) is

$$p_1 = \pm \frac{13\sqrt{21}}{31} - \frac{143}{93}i, \tag{33}$$

$$\alpha = \pm \frac{2\sqrt{21}}{3}, \quad \beta = -\frac{2}{3}, \tag{34}$$

$$\sigma_1 = \sigma_2 = \frac{53}{27}, \quad \omega = -\frac{13k^2}{2}, \quad |h|^2 = -\frac{39k^2}{4q_i}. \tag{35}$$

The validity of these solutions is verified by the direct substitution into the underlying eqs. (5), (6).

*Example C*

As an example of CGLEs with three free parameters ($p_r$, $q_i$, $k^2$), we take eqs. (22 – 28) with $p_2$ purely real, i.e.

$$p_2 = p_r \text{ (real)}, \qquad q_1 = q_2 = iq_i, \tag{36}$$

$$\alpha = \sqrt{14}, \quad \beta = \mp\sqrt{2}, \quad \sigma_1 = \sigma_2 = 3, \tag{37}$$

$$\omega = \pm \frac{p_r k^2}{\sqrt{2}}, \quad p_1 = \pm \frac{(-\sqrt{14}+4i)p_r}{15\sqrt{2}} = p_{1r} + ip_{1i}, \tag{38}$$

$$|h|^2 = \pm \frac{3 p_r k^2}{2\sqrt{2} q_i}, \qquad |g|^2 = \pm \frac{9 p_r k^2}{2\sqrt{2} q_i}, \tag{39}$$

$$\Omega_1 = \pm \frac{\sqrt{14} p_r k^2}{15\sqrt{2}}, \quad \gamma_1 = \pm \frac{23 p_r k^2}{30\sqrt{2}}, \quad \Omega_2 = 0, \quad \gamma_2 = \pm \frac{3 p_r k^2}{2\sqrt{2}}, \tag{40}$$

and the merit of eqs. (36 – 40) is an elegant simplification of system (5, 6). The amplitude functions

$$A = g \exp\left(\mp i \frac{\sqrt{14} p_r k^2}{15\sqrt{2}} t\right) \frac{\exp(kx \mp p_r k^2 t/\sqrt{2})}{[1+\exp(kx \mp p_r k^2 t/\sqrt{2})]^{1+i\sqrt{14}}}, \tag{41}$$

$$B = \frac{h}{[1+\exp(kx \mp p_r k^2 t/\sqrt{2})]^{1\mp i\sqrt{2}}}, \tag{42}$$

solve the coupled CGLEs,

$$iA_t \pm \frac{(-\sqrt{14}+4i)p_r}{15\sqrt{2}} A_{xx} + iq_i(|A|^2+|B|^2)A = \pm i\frac{23 p_r k^2}{30\sqrt{2}} A, \tag{43}$$

$$iB_t + p_r B_{xx} + iq_i(|A|^2+|B|^2)B = \pm i\frac{3 p_r k^2}{2\sqrt{2}} B, \tag{44}$$

with signs of $p_r$ and $q_i$ taken to make the right hand side of eq. (39) positive.

Expressions (41, 42) constitute an exact solution of two coupled, nonlinear partial differential eqs. (43, 44) with three arbitrary parameters ($k^2$, $p_r$ and $q_i$). Regarding the choice of signs in the symbol '±', either the upper or the lower sign must be taken throughout the entire set of eqs. (36 – 44) in a consistent manner.

For the '+' sign in eqs. (43, 44), if $p_r < 0$, linear damping is present in both components, and diffusion spreading will occur in component $A$. These two factors will contribute to the attenuation of the wave envelopes. However, eq. (39) will dictate that $q_i < 0$, which implies the existence of a cubic gain, and this will sustain the front.

Examples of front patterns propagating to the right and left are shown in Figures 1 and 2 respectively. The intensities of the plane-wave background supporting the fronts, which are $|g|^2$ and $|h|^2$ in the present case, depend on the precise structure of the solution. For solutions eqs. (36 – 42) of eqs. (43), (44), $|A|^2$ is generally larger than $|B|^2$, and the difference is more profound for large $k$ (Figure 3).

## 5. The second family of exact solutions (purely real $q_1$, $q_2$)

Another family of exact solutions is obtained by taking purely real $q_1$, $q_2$:

$$q_2 = \varepsilon \sqrt{\frac{\sigma_2}{\sigma_1}} q_1 = \varepsilon \sqrt{\frac{\sigma_2}{\sigma_1}} q_r, \qquad q_i = 0, \qquad \varepsilon = \pm 1 \ . \tag{45}$$

In this case, a solution for eqs. (14 – 18) with four degrees of freedom, i.e., arbitrary $q_r$, $p_{2r}$, $p_{2i}$ (relation (22) still being valid) and $k^2$, is

$$\beta^2 - \frac{3 p_{2r} \beta}{p_{2i}} - 2 = 0, \tag{46}$$

$$5\alpha^4 + \left( \frac{81 p_{2r}^2}{p_{2i}^2} - \frac{54 \beta p_{2r}}{p_{2i}} - 8 \right) \alpha^2 - 4 = 0, \tag{47}$$

$$\sigma_1 = -\frac{5\alpha^2 + 2}{3(\alpha^2 - 2)}, \qquad \sigma_2 = -\frac{3(\beta^2 + 2)}{\beta^2 - 2}, \tag{48}$$

$$\omega = k^2 (\beta p_{2r} + p_{2i}) \qquad |h|^2 = -\frac{3 \omega p_{2r}}{2 q_r p_{2i}}, \tag{49}$$

$$p_1 = -\frac{3 \varepsilon \omega p_{2r} (\sigma_1 + 1)[\alpha^2 - 2 + 3\alpha i]}{2 p_{2i} k^2 (\alpha^2 + 1)(\alpha^2 + 4)} \sqrt{\frac{\sigma_2}{\sigma_1}} . \tag{50}$$

Again the parameters must be chosen so as to make real solutions possible, e.g., the quadratic equations written above for β and $\alpha^2$ must not lead to complex roots. In this case, eqs. (7), (8), (11), (19), and (45 – 50) furnish an exact, analytical solution for the coupled-CGLE system, eqs. (5), (6).

As a simple numerical example, consider $q_1 = q_r$, where $q_r < 0$ (arbitrary) and $p_2 = 1 + 3i$, we obtain

$$q_2 = \pm \frac{9\sqrt{5}}{5} q_r, \qquad p_1 = \pm\sqrt{5} + \frac{5}{3} i, \tag{51}$$

$$\alpha = \mp\frac{\sqrt{5}}{5}, \quad \beta = -1, \quad \sigma_1 = \frac{5}{9}, \quad \sigma_2 = 9, \tag{52}$$

$$\omega = 2k^2, \quad |h|^2 = -\frac{k^2}{q_r}, \tag{53}$$

and with $g$, $\Omega_1$, $\Omega_2$, $\gamma_1$, $\gamma_2$ given by eq. (19). Calculations similar to those presented in Section 4 can be performed, but will not be pursued here.

## 6. Stability of domain walls

The stability of wave profiles is of crucial importance, since it determines if such patterns can be observed in an experiment. The stability of domain walls was studied by numerical simulations of perturbed wave profiles. The spatial derivative in $x$ in eqs. (5), (6) was approximated by a Crank–Nicholson scheme, i.e. a semi–implicit, second-order central difference operator. The time derivative was handled by means of a simple forward Euler operation. The typical number of grid points in the spatial domain was around 2000. The time step was adjusted until consistent results were obtained when the number of temporal grid points doubled. The linear and nonlinear gain/loss was treated explicitly.

As the number of parameters in eqs. (5, 6) is vast (complex $p_1$, $p_2$, $q_1$, $q_2$, real $\gamma_1$, $\gamma_2$), we shall demonstrate some simple examples of stability versus instability. As a typical case, we choose the first family of exact solutions of eqs. (5), (6), as given by eqs. (41), (42). With $p_r = 1$, $q_i = 1$, $k = 0.01$, and the positive sign for $\omega$ taken in eq. (38), random amplitude disturbance of 1% was imposed on the fronts. Figure 4 shows that the initially perturbed patterns can persist for a reasonable amount of time.

With $p_r = 1$, $q_i = -1$, $k = 0.1$, and the negative sign for $\omega$ taken in eq. (38), random amplitude disturbance of 1% was again imposed on the fronts. In sharp contrast with the previous case, apparently exponential growth is observed, starting around $t = 30$ (Figure 5). For this particular choice of the parameters, stability is more likely attained for smaller values of $k$.

To verify the numerical simulations, as well as to provide a deeper insight of the underlying physics, an order-of-magnitude balance was examined too. If one considers eq. (5) at the onset of the exponential growth, one can simplify the dynamics through the following assumptions:

● the term $|B|^2$ is nearly zero there (while $|A|$ corresponds to a nonzero plane wave, as required by the definition of the 'domain wall');

● the term $A_{xx}$ can be neglected as the wave profile is nearly flat there.

As such the dynamics of the growth ('imaginary part' of (5)) is governed by

$$A_t = -q_i |A|^2 A + \gamma_1 A, \qquad (54)$$

and thus the right-hand side of (54) will provide one estimate of the 'time derivative of $A$', which we shall call the 'theoretical growth rate' here.

In terms of numerical simulations, if superscripts denote the discretized time, a simple forward Euler scheme will provide a leading order approximation of the time derivative:

$$(A^{n+1} - A^n)/\Delta t \,. \tag{55}$$

From the numerical data obtained in the simulations, we compute (55) directly and term this quantity the 'numerical time derivative' of *A*. Figure 6 shows the comparison between the 'theoretical growth rate' versus this 'numerical time derivative' at a typical spatial location ($x = 260$). The agreement is remarkable. Thus we conclude that the numerical simulations provide a very reasonable description of the nonlinear dynamics, and simple scenarios for stability and instability of 'domain walls' have been demonstrated.

## 7. Conclusions

In this work, we have presented exact analytical solutions for domain walls, i.e., pairs of mutually locked fronts with opposite polarities, in a system of nonlinearly coupled CGLEs (complex Ginzburg-Landau equations). Due to the presence of amplification and attenuation, the analysis of CGLEs is substantially more involved than the energy conserving nonlinear Schrödinger equation.[30–32]

The efficiency of the Bekki-Nozaki modified Hirota bilinear operator in solving such systems has been demonstrated before in various settings, such as inhomogeneous media which correspond to CGLEs with variable coefficients,[33] and interactions of solitary pulses and fronts.[34] Here we have focused on configurations with fronts featuring opposite polarities in both components of the CGLE system. The equations for the wave profiles were solved by means of computer-assisted algebraic manipulations.

Exact solutions have been obtained for special cases where either

● the cross- and self-phase modulation terms are absent (purely imaginary $q_1 = q_2$ in eqs. (5), (6), which implies that the nonlinearity is dissipative), or

● the opposite case when cubic amplification/dissipation is absent (real $q_1 = q_2$).

Several aspects of the present analysis can be further enhanced. More general exact solutions may be feasible with other combinations of parameters, e.g., purely imaginary $p_1$, $p_2$ will yield differential operators of the reaction-diffusion type. An obviously important issue which calls for additional analysis is the modulation stability of the plane–wave background.[35] A related issue is the numerical simulation of the subsequent development of any possible modulation instability. Bifurcations and symmetries of the front patterns can also be investigated in future studies.[36]

The scheme for the generation of exact solutions for 'localized pulse – front' and 'fronts of opposite polarities' complexes can also be extended to other two-component nonlinear evolution equations with complex coefficients,[37, 38] and would be promising in the applications to science and engineering disciplines.[39, 40]

## Acknowledgements

Partial financial support has been provided by the Research Grants Council of Hong Kong through contract HKU 7120/08E.

## Appendix

Some details on the derivation of eqs. (14 – 18) are now given. From elementary calculus one can readily establish (*G*, *f* are complex functions, *m* is a complex number):

$$D_t\left(\frac{G}{f^m}\right) = \frac{D_{m,t}G \cdot f}{f^{m+1}}, \tag{A1}$$

$$D_x^2\left(\frac{G}{f^m}\right) = \frac{D_{m,x}^2 G \cdot f}{f^{m+1}} - \frac{m(m+1)}{2}\left(\frac{D_x^2 f \cdot f}{f^{m+2}}\right). \tag{A2}$$

Hence with eqs. (5 – 7), one can deduce the validity of eqs. (9, 10). The crucial argument of this paper is the factorization process eqs. (12, 13). It is sufficient to illustrate the details for one of the CGL equation system eqs. (9, 10), say eq. (9), as calculations for the other component are similar. If *f* is given by eq. (11), the ordinary Hirota derivative can be simplified through identity eq. (4) as well. One now equates the constant term, coefficients of exp[$kx - \omega t$] and exp[$2(kx - \omega t)$] of eq. (12), and deduces that

$$q_1|g|^2 = \sigma_1 q_2|h|^2,$$

$$-p_1 m(m+1)k^2 = (1 + \sigma_1)q_2|h|^2.$$

The second expression above is eq. (16) in the main text.

Applying eq. (12) to eq. (9), one factor of $f$ can be cancelled throughout the equation, and we have

$$[iD_{m,t} + p_1 D_{m,x}^2 + \Omega_1 - i\gamma_1]G \cdot f + Gq_2|h|^2[1 + \sigma_1 \exp(kx - \omega t)] = 0,$$

which is a bilinear equation. Using the basic principles in simplifying these modified Hirota derivatives of exponential functions as described in eq. (4), one now arrives, on considering the coefficients of exp[$kx - \omega t$] and exp[$2(kx - \omega t)$],

$$i\omega + p_1 k^2(m-2) + \frac{q_2 h h^*(\sigma_1 - 1)}{m} = 0, \tag{A3}$$

$$-i\omega + p_1 k^2 + \Omega_1 - i\gamma_1 + q_2 h h^* = 0.$$

Using eq. (16) to eliminate $p_1$ in eq. (A3) will produce equation eq. (14) of the main text. The other members in eqs. (14 – 18) are derived in a similar manner.

## References

1) A. D. D. Craik: *Wave Interactions and Fluid Flows* (Cambridge University Press, Cambridge, 1984).

2) M. C. Cross and P. C. Hohenberg: Rev. Mod. Phys. **65** (1993) 851.

3) F. T. Arecchi, S. Boccaletti and P. L. Ramazza: Phys. Rep. **318** (1999) 1.

4) M. Ipsen, L. Kramer and P. G. Sorensen: Phys. Rep. **337** (2000) 193.

5) I. S. Aranson and L. Kramer: Rev. Mod. Phys. **74** (2002) 99.

6) Y. S. Kivshar and G. P. Agrawal: *Optical Solitons: From Fibers to Photonic Crystals* (Academic Press, San Diego, 2003).

7) B. A. Malomed: 'Complex Ginzburg-Landau Equation', in *Encyclopedia of Nonlinear Science*, Edited by A. Scott, p. 157 (Routledge, New York, 2005).

8) B. A. Malomed: *Soliton Management in Periodic Systems* (Springer, New York, 2006).

9) N. Akhmediev and A. Ankiewicz: 'Solitons of the Complex Ginzburg-Landau Equation' in *Spatial Solitons*, Edited by S. Trillo, p. 311 (Springer, New York, 2002).

10) Y. Matsuno: *The Bilinear Transformation Method* (Academic Press, New York, 1984).

11) K. Nozaki and N. Bekki: J. Phys. Soc. Jpn. **53** (1984) 1581.

12) H. Herrero and H. Riecke: Physica D **85** (1995) 79.

13) C. Hemming and R. Kapral: Physica D **168** (2002) 10.

14) W. van Saarloos: Phys. Rep. **386** (2003) 29.

15) M. J. Smith and J. A. Sherratt: Phys. Rev. E **80** (2009) 046209.

16) P. Gutiérrez, D. Escaff and O. Descalzi: Phil. Trans. Roy. Soc. A **367** (2009) 3227.

17) J. Atai and B. A. Malomed: Phys. Lett. A **246** (1998) 412.

18) H. Sakaguchi and B. A. Malomed: Physica D **154** (2001) 229.

19) B. A. Malomed: Chaos **17** (2007) 037117.

20) W. J. Firth and P. V. Paulau: Eur. Phys. J. D **59** (2010) 13.

21) M. C. Cross: Phys. Rev. Lett. **57** (1986) 2935.

22) B. A. Malomed, A. A. Nepomnyashchy and M. I. Tribelsky: Phys. Rev. A **42** (1990) 7244.

23) B. A. Malomed: Phys. Rev. E **50** (1994) R3310.

24) B. A. Malomed: Physica Scripta **57** (1997) 115.

25) W. Hong: Opt. Comm. **281** (2008) 6112.

26) Y. Yoshida and Y. Kimura: J. Phys. Soc. Jpn. **78** (2009) 084801.

27) B. Bruhn: Phys. Plasmas **13** (2006) 023505.

28) N. Bekki and K. Nozaki: Phys. Lett. A **110** (1985) 133.

29) R. Radhakrishnan and M. Lakshmanan: J. Phys. A **28** (1995) 2683.

30) Z. Y. Huang, S. Jin, P. A. Markowich and C. Sparber: Wave Motion **46** (2009) 15.

31) B. F. Feng and T. Kawahara: Wave Motion **45** (2007) 68.

32) A. B. Aceves: Wave Motion **45** (2007) 48.

33) O. S. Pak, K. W. Chow, C. K. Lam, K. Nakkeeran and B. A. Malomed: J. Phys. Soc. Jpn. **78** (2009) 084001.

34) T. L. Yee and K. W. Chow: J. Phys. Soc. Jpn. **79** (2010) 124003.

35) H. Sakaguchi: Prog. Theor. Phys. **85** (1991) 417.

36) H. Sakaguchi and H. Takeshita: J. Phys. Soc. Jpn. **77** (2008) 054003.

37) E. Yomba, T. C. Kofane, F. B. Pelap: J. Phys. Soc. Jpn. **65** (1996) 2337.

38) K. Matsuba, K. Imai and K. Nozaki: J. Phys. Soc. Jpn. **66** (1997) 1668.

39) A. A. Golovin, A. A. Nepomnyashchy and B. J. Matkowsky: Physica D **160** (2001) 1.

40) A. A. Izquierdo, M. A. G. Leon and J. M. Guilarte: Phys. Rev. D **65** (2002) 085012.

**Figures Captions**

(1) Figure 1: Fronts moving to the right (eqs. (41, 42), $p_r = 2$, $q_i = 1$, $k = 1$, with the positive sign for $\omega$ in eq. (38)), Top: $|A|^2$ versus $x$ and $t$, Bottom: $|B|^2$ versus $x$ and $t$.

(2) Figure 2: Fronts moving to the left (eqs. (41, 42), $p_r = 2$, $q_i = -1$, $k = 1$, with the negative sign for $\omega$ in eq. (38)), Top: $|A|^2$ versus $x$ and $t$, Bottom: $|B|^2$ versus $x$ and $t$.

(3) Figure 3: (Color Online) Intensities $|A|^2$, $|B|^2$ of eqs. (41, 42) versus $k$, for $p_r = 1$, $|q_i| = 1$.

(4) Figure 4: Numerical simulation showing a *stable* evolution of perturbed fronts given by eqs. (41, 42), with random disturbance of 1% amplitude being imposed on the fronts ($p_r = 1$, $q_i = 1$, $k = 0.01$, with the positive sign for $\omega$ taken in eq. (38)).

(5) Figure 5: Numerical simulation showing an *unstable* evolution of perturbed front $A$ given by eqs. (41, 42), with random disturbance of 1% amplitude being imposed on the fronts ($p_r = 1$, $q_i = -1$, $k = 0.1$, with the negative sign for $\omega$ taken in eq. (38)).

(6) Figure 6: (Color Online) Comparison between the theoretical growth rate, eq. (54), and the numerical time derivative, eq. (55), versus time $t$ for the unstable front $A$ at a typical point, $x = 260$ (solid line: theoretical growth rate, dashed line: numerical time derivative of $A$).

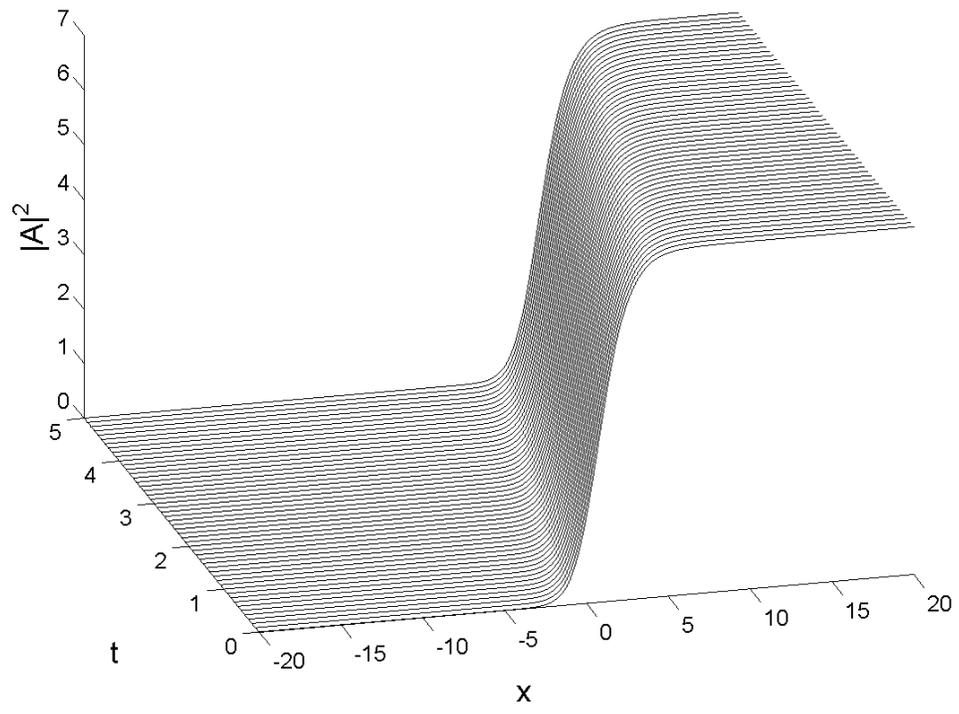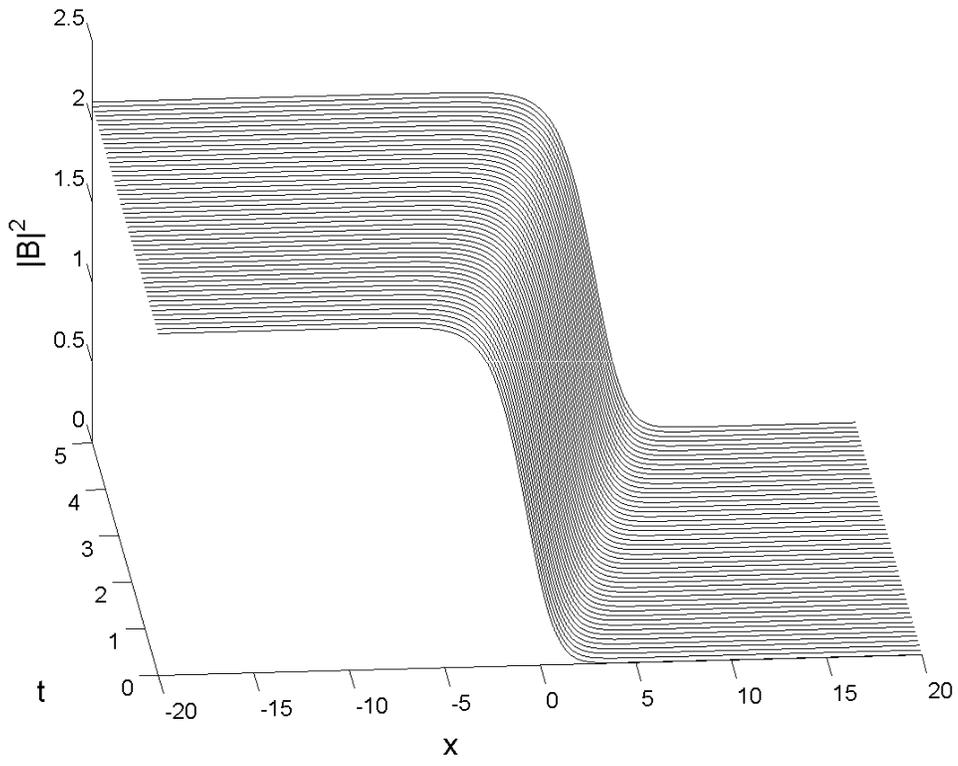

**Figure 1**

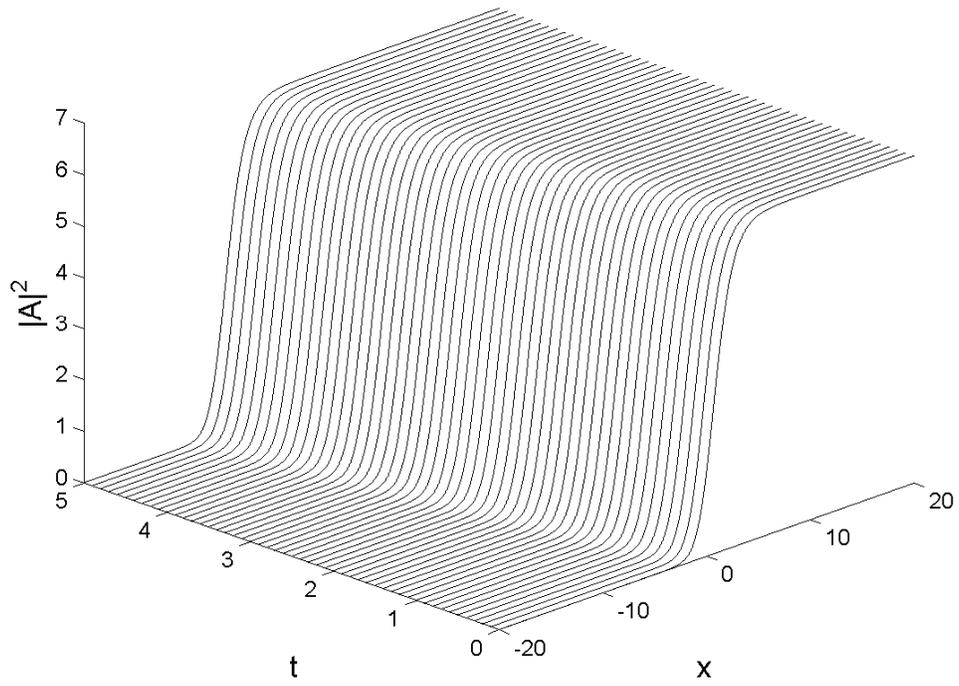

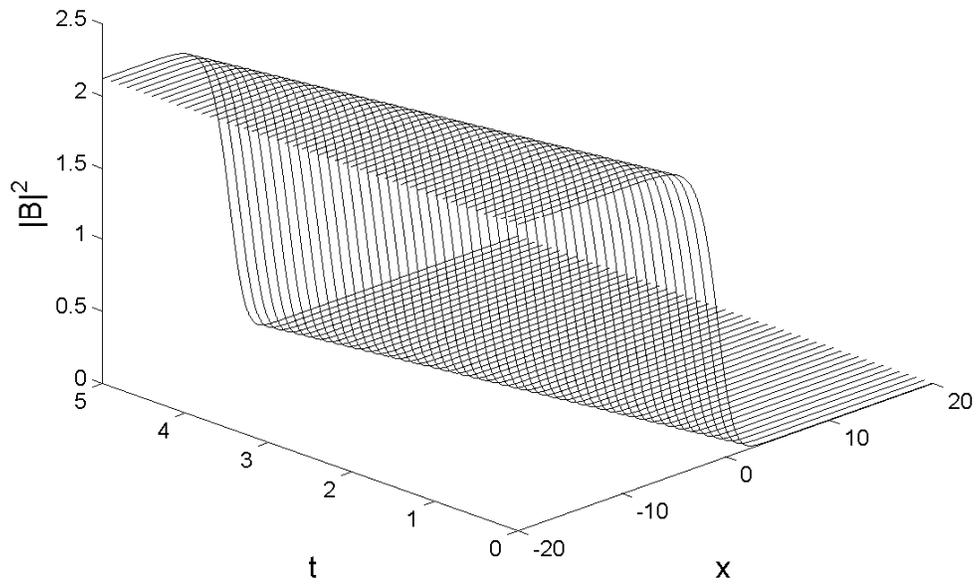

**Figure 2**

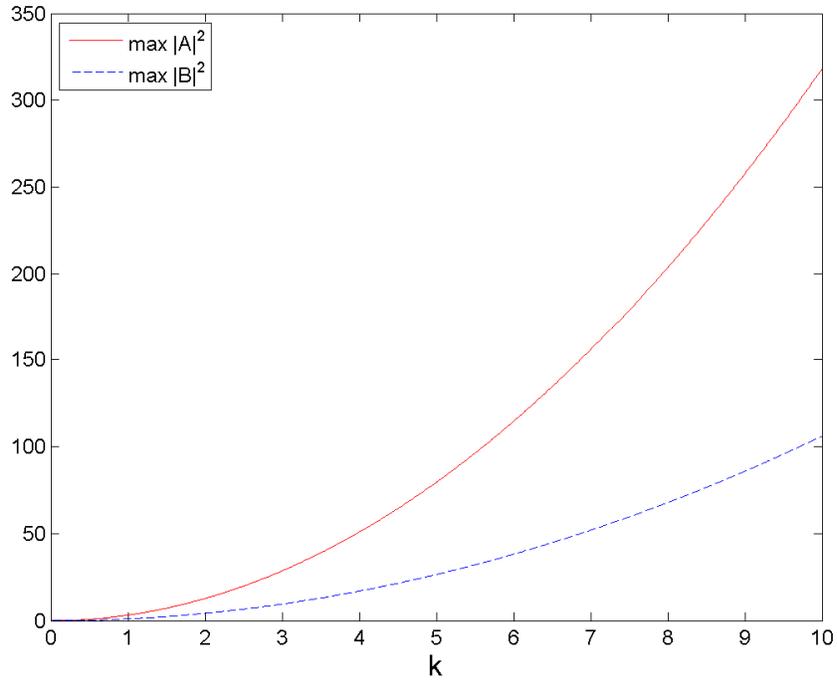

**Figure 3**

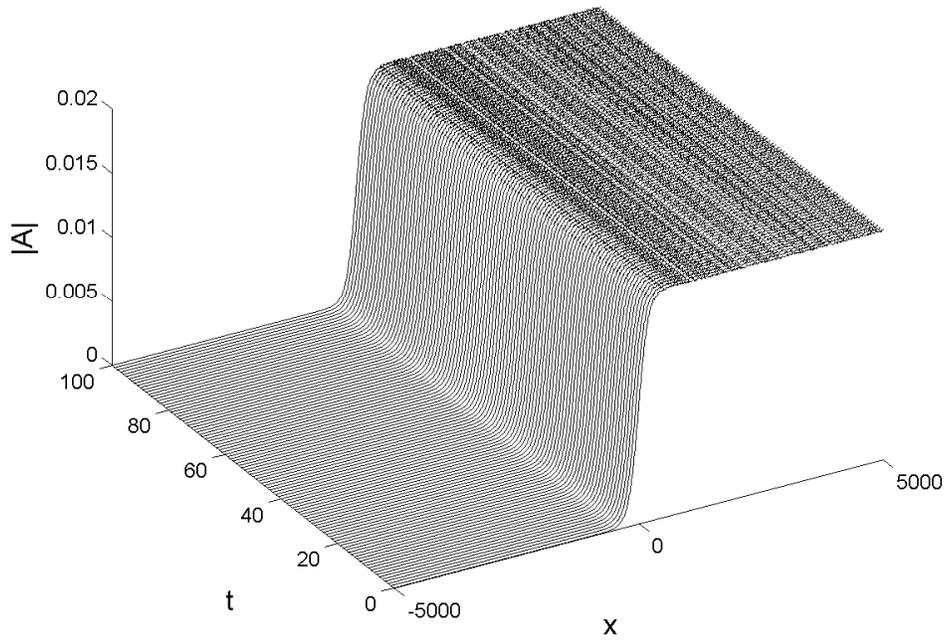

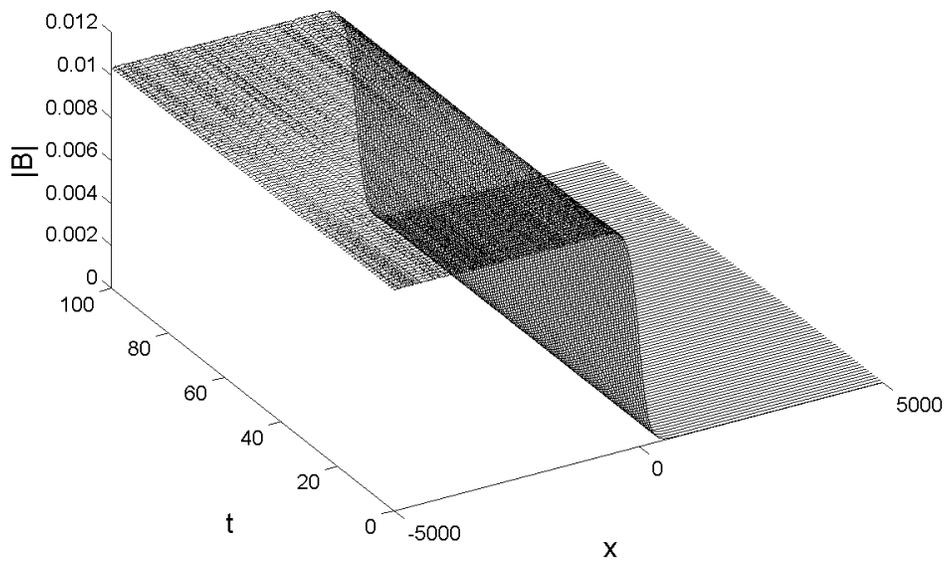

**Figure 4**

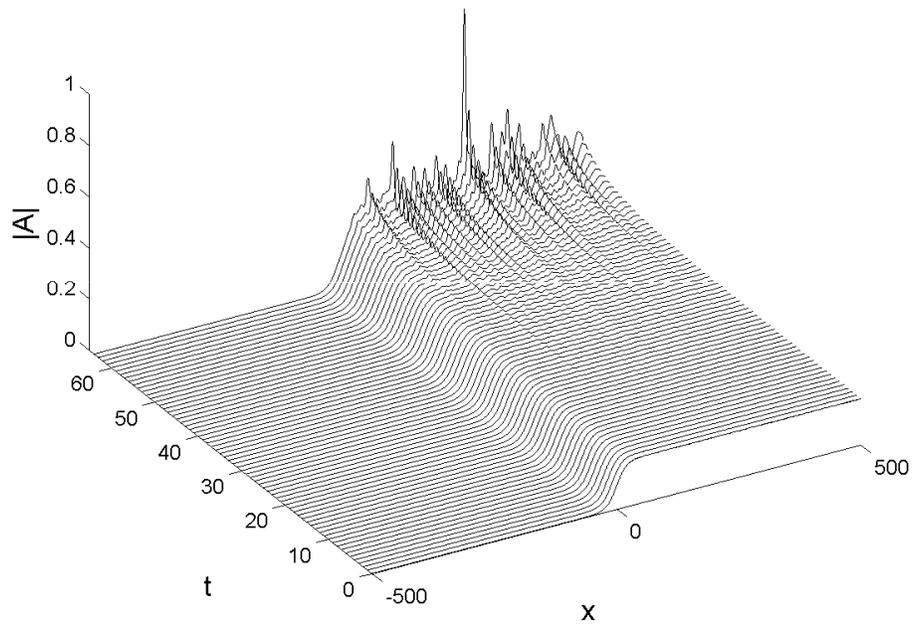

**Figure 5**

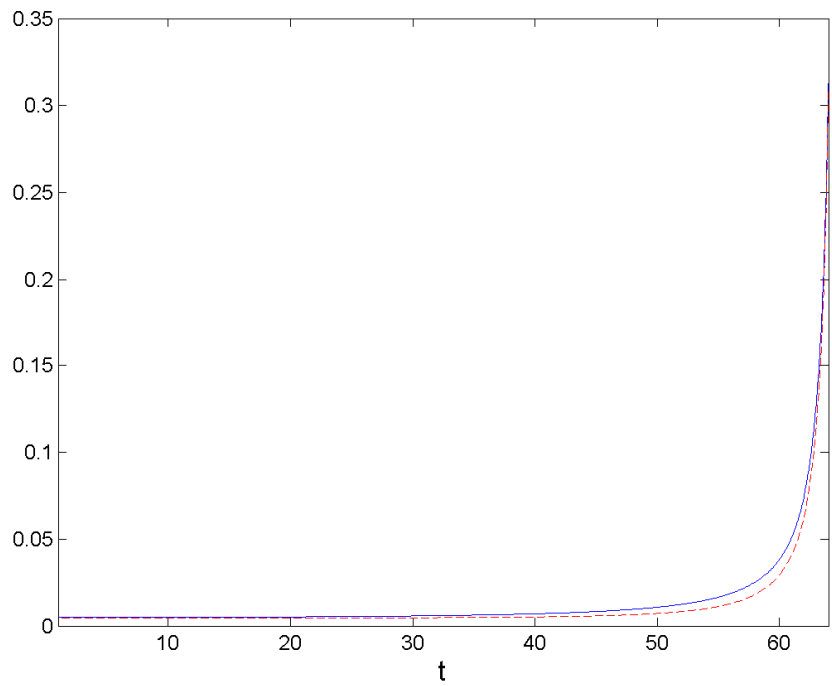

**Figure 6**